\documentclass{JAC2003}
\usepackage{graphicx}
\usepackage{booktabs}
\newtheorem{proposition}{Proposition}

\begin{document}
\title{TWISS PARAMETERS OF COUPLED PARTICLE BEAMS\\ WITH EQUAL EIGENEMITTANCES
\vspace{-0.5cm}}

\author{V.Balandin\thanks{vladimir.balandin@desy.de}, 
R.Brinkmann, W.Decking, N.Golubeva \\
DESY, Hamburg, Germany}

\maketitle

\begin{abstract}
We show that the 1D Courant-Snyder theory can be considered as a partial case of 
the multidimensional theory of coupled particle beams with equal eigenemittances.
\end{abstract}

\vspace{-0.2cm}
\section{INTRODUCTION}

\vspace{-0.1cm}
The parametrization of coupled beam motion has been studied intensively over the
past decades. Nevertheless, there is still no representation of general coupled 
motion that would be as complete and as widely accepted as the 1D Courant-Snyder 
theory. But is it really so, that the Courant-Snyder theory is, essentially, 
the theory which is applicable only to the 1D linear motion? 
Our answer is that the one-dimensionality of motion is not the main source of the 
elegance of the Courant-Snyder approach. It is the property of the beam matrix 
to be proportional to the matrix which is simultaneously symmetric positive definite 
and symplectic. Because this property is independent from the phase space dimensions 
and is the characteristic property of the particle beams with equal eigenemittances, 
the purpose of this paper is to show that the 1D Courant-Snyder theory can be considered 
as a natural partial case of the multidimensional theory of coupled particle beams 
with equal eigenemittances. It is also possible to say (in the opposite way), that in 
this paper we extend 
the 1D Courant-Snyder formalism to the multidimensional theory of beams with equal 
eigenemittances. Due to space limitation, we consider only the most basic theoretical
questions and the more detailed study will be presented in a separate publication.

\vspace{-0.2cm}
\section{BEAM MATRIX AND EIGENEMITTANCES}

\vspace{-0.1cm}
Let us consider a collection of points in $2n$-dimensional 
phase space (a particle beam) and let, for each particle, 
 
\vspace{-0.2cm}
\noindent
\begin{eqnarray}
z \,=\,(q, p)^{\top}\,=\,
(q_1, \ldots, q_n, p_1, \ldots,p_n)^{\top}
\label{intr_1}
\end{eqnarray}

\vspace{-0.2cm}
\noindent
be a vector of canonical coordinates and momenta. Then, as usual, 
the beam (covariance) matrix is defined as

\vspace{-0.2cm}
\noindent
\begin{eqnarray}
\Sigma =
\Big<
\big(z - \big< z \big> \big) 
\cdot 
\big(z - \big< z \big> \big)^{\top}
\Big>
\stackrel{\mbox{\tiny def}}{=}
\left(
\begin{array}{ll}
\Sigma_{qq} & \Sigma_{qp}\\
\Sigma_{pq} & \Sigma_{pp}\\
\end{array}
\right),
\label{vectZ_1}
\end{eqnarray}

\vspace{-0.2cm}
\noindent
where the brackets $\,\big< \,\cdot \,\big>\,$ denote an average over 
a distribution of the particles in the beam and $\,n \times n\,$ submatrices 
of the $\,2n \times 2n\,$ matrix $\Sigma$ satisfy 

\vspace{-0.2cm}
\noindent
\begin{eqnarray}
\Sigma_{qq} \,=\, \Sigma_{qq}^{\top},
\;\;\;\;
\Sigma_{pp} \,=\, \Sigma_{pp}^{\top},
\;\;\;\;
\Sigma_{pq} \,=\, \Sigma_{qp}^{\top}.
\label{vectZ_3}
\end{eqnarray}

\vspace{-0.2cm}
\noindent
One says that the beam is uncoupled if all four submatrices of the matrix 
$\Sigma$ are diagonal matrices. By definition, the matrix $\Sigma$ is symmetric 
positive semidefinite and in the following we will restrict our considerations 
to the situation when this matrix is nondegenerated and therefore positive definite. 
For simplification of notations and without loss of generality, we will also assume 
that the beam has vanishing first-order moments, i.e. $\big< z \big> \,=\, 0$.

Let $s$ be the independent variable (time or path length along design orbit) and 
let $\,M(\tau)\,$ be the linear transfer matrix which propagates particle coordinates
from the state $\,s = 0\,$ to the state $\,s = \tau$, i.e let

\vspace{-0.2cm}
\noindent
\begin{eqnarray}
z(\tau) \,=\, M(\tau)\, z(0).
\label{intr_4}
\end{eqnarray}

\vspace{-0.2cm}
\noindent
Then the beam matrix $\Sigma$ evolves between these two
states according to the rule

\vspace{-0.2cm}
\noindent
\begin{eqnarray}
\Sigma(\tau) \,=\, M(\tau)\, \Sigma(0)\, M^{\top}(\tau).
\label{intr_5}
\end{eqnarray}

\vspace{-0.2cm}
\noindent
If the matrix $M(\tau)$ is symplectic and satisfies the relations

\vspace{-0.2cm}
\noindent
\begin{eqnarray}
M^{\top}(\tau)\,J M(\tau)\,=\,M(\tau)\,J M^{\top}(\tau)\,=\,J,
\label{vectZ_6}
\end{eqnarray}

\vspace{-0.2cm}
\noindent
where

\vspace{-0.4cm}
\noindent
\begin{eqnarray}
J \;=\;
\left(
\begin{array}{rr}
0 & I\\
-I & 0
\end{array}
\right)
\label{matr_J}
\end{eqnarray}

\vspace{-0.2cm}
\noindent
is the $2n \times 2n$ symplectic unit matrix and $I$ is the $n \times n$
identity matrix, then the transport equation (\ref{intr_5}) can be transformed
into the following equivalent form

\vspace{-0.2cm}
\noindent
\begin{eqnarray}
(\Sigma J) (\tau)\;=\; M(\tau)\cdot (\Sigma J)(0)\cdot M^{-1}(\tau).
\label{vectZ_7}
\end{eqnarray}

\vspace{-0.2cm}
\noindent
From this form of the equation (\ref{intr_5}) we see that the eigenvalues 
of the matrix $\Sigma J$ are invariants, because (\ref{vectZ_7}) is a similarity 
transformation. The matrix $\Sigma J$ is nondegenerated and is similar to 
the skew symmetric matrix $\,\Sigma^{1/2} J \,\Sigma^{1/2}$

\vspace{-0.2cm}
\noindent
\begin{eqnarray}
\Sigma J\;=\; \Sigma^{1/2} \cdot 
(\Sigma^{1/2} J \,\Sigma^{1/2}) \cdot \Sigma^{-1/2},
\label{vectZ_8}
\end{eqnarray}

\vspace{-0.2cm}
\noindent
which means that its spectrum is of the form

\vspace{-0.2cm}
\noindent
\begin{eqnarray}
\pm i \epsilon_1, \,\ldots, \,\pm i \epsilon_n,
\label{vectZ_9}
\end{eqnarray}

\vspace{-0.2cm}
\noindent
where all $\epsilon_m > 0$ and $i$ is the imaginary unit.

The quantities $\epsilon_m$ are called eigenemittances 
and are generalizations of the usual 1D rms emittances

\vspace{-0.2cm}
\noindent
\begin{eqnarray}
\varepsilon_m 
= \sqrt{\big<q_m^2\big>\big<p_m^2\big> -\big<q_m p_m\big>^2}
\label{intr_12}
\end{eqnarray}

\vspace{-0.2cm}
\noindent
to the fully coupled case ~\cite{Neri}.
It is not difficult to prove that the set of all eigenemittances
$\{\epsilon_m\}$ and the set of all projected emittances $\{\varepsilon_m\}$
of a given beam matrix $\Sigma$ coincide if and only if the beam is uncoupled. 

The other approach to the concept of eigenemittances is the way pointed out 
by Williamson's theorem (see, for example, references in ~\cite{Neri}).
This theorem tells us that one can diagonalize any positive definite 
symmetric matrix $\Sigma$ by congruence using a symplectic matrix $T$

\vspace{-0.2cm}
\noindent
\begin{eqnarray}
T \,\Sigma \, T^{\top} \;=\; D,
\label{A2_1}
\end{eqnarray}

\vspace{-0.2cm}
\noindent
and that the diagonal matrix $D$ has the very simple form

\vspace{-0.2cm}
\noindent
\begin{eqnarray}
D\,=\,\mbox{diag}(\Lambda, \Lambda),
\;\;
\Lambda \,=\, \mbox{diag} (\epsilon_1, \, \ldots, \, \epsilon_ n) > 0,
\label{A2_2}
\end{eqnarray}

\vspace{-0.2cm}
\noindent
where the diagonal elements $\epsilon_m$ are the moduli of the eigenvalues 
of the matrix $\Sigma J$. The matrix $T$ in (\ref{A2_1}) is not unique, but 
the diagonal entries of the Williamson's normal form $D$ (eigenemittances) 
are unique up to a reordering.

\section{TWISS PARAMETERS\\ OF THE BEAM MATRIX WITH EQUAL EIGENEMITTANCES}

\vspace{-0.1cm}
Let us assume that the beam matrix $\Sigma$ has all eigenemittances 
equal to each other and equal to the value $\epsilon > 0$. Then, according to 
the Williamson's theorem, there exists a symplectic matrix $T$ such that

\vspace{-0.2cm}
\noindent
\begin{eqnarray}
\Sigma \, = \, \epsilon \, T^{-1} \, T^{-\top}
\,\stackrel{\mbox{\tiny def}}{=}\, \epsilon \,W.
\label{matr_S2}
\end{eqnarray}

\vspace{-0.2cm}
\noindent
The matrix $W$ in (\ref{matr_S2}) is independent from any 
particular choice of the diagonalizing matrix $T$ in (\ref{A2_1}) and is 
simultaneously symmetric positive definite and symplectic.
We will call it the Twiss matrix and will parametrize it as follows

\vspace{-0.4cm}
\noindent
\begin{eqnarray}
W \;=\;
\left(
\begin{array}{cr}
\beta & - \alpha\\
-\alpha^{\top} & \gamma
\end{array}
\right),
\label{matr_S}
\end{eqnarray}

\vspace{-0.2cm}
\noindent
where the $n \times n$ submatrices
$\beta = \beta^{\top}$, $\alpha$ and $\gamma = \gamma^{\top}$ are the natural
matrix generalizations of the corresponding 1D scalar Twiss parameters.
Due to symplecticity of the matrix $W$ the matrix Twiss parameters satisfy 
the relations

\vspace{-0.2cm}
\noindent
\begin{eqnarray}
\beta \, \gamma \;=\; I \,+\,\alpha^2,
\label{betas_4}
\end{eqnarray}

\vspace{-0.4cm}
\noindent
\begin{eqnarray}
\alpha \, \beta \;=\;\beta\,\alpha^{\top},
\label{betas_5}
\end{eqnarray}

\vspace{-0.4cm}
\noindent
\begin{eqnarray}
\gamma \,\alpha  \;=\;\alpha^{\top}\,\gamma.
\label{betas_7}
\end{eqnarray}

\vspace{-0.2cm}
\noindent
Because the matrix $W$ is positive definite, both its submatrices
$\beta$ and $\gamma$ are also positive definite and, therefore, nondegenerated.
From this it follows that not all relations (\ref{betas_4})-(\ref{betas_7})
are independent. For example, the relations (\ref{betas_5}) or 
the relations (\ref{betas_7}) can be omitted.

The inverse of the Twiss matrix $W$ is given by the formula

\vspace{-0.2cm}
\noindent
\begin{eqnarray}
W^{-1} \;=\;
\left(
\begin{array}{ll}
\gamma &  \alpha^{\top}\\
\alpha & \beta
\end{array}
\right),
\label{matr_S_M1}
\end{eqnarray}

\vspace{-0.2cm}
\noindent
and, therefore, the natural multidimensional analogy of 
the positive definite 1D Courant-Snyder invariant 

\vspace{-0.2cm}
\noindent
\begin{eqnarray}
I_{cs}\,=\,z^{\top} W^{-1} z  
\label{matr_S_M19}
\end{eqnarray}

\vspace{-0.2cm}
\noindent
can be written as follows

\vspace{-0.2cm}
\noindent
\begin{eqnarray}
I_{cs}  = 
q^{\top} \gamma \,q +
q^{\top}\alpha^{\top} p +
p^{\top} \alpha \,q +
p^{\top} \beta \,p.
\label{betas_77}
\end{eqnarray}

\vspace{-0.2cm}
In general, the eigenemittances of the matrix $\Sigma$ have to be found as
solution of the eigenvalue problem for the matrix $\Sigma J$. But if one wants 
only to know if all of them are equal to each other or not, then the solution of 
the eigenvalue problem is not necessary. It can be done by simple matrix 
multiplication as explained in the following proposition.

\vspace{-0.2cm}
\begin{proposition}

The beam matrix $\Sigma$ has all eigenemittances equal to each other 
and equal to the value $\epsilon > 0$ if and only if the equality 

\vspace{-0.2cm}
\noindent
\begin{eqnarray}
( \Sigma \, J )^2 \, + \, \epsilon^2 \, I \, = \, 0
\label{CC_4}
\end{eqnarray}

\vspace{-0.2cm}
\noindent
holds, i.e. if and only if the matrix $(\Sigma \, J)^2$
is a negative scalar matrix.

\end{proposition}

\section{PARAMETRIZATION OF BEAM TRANSFER MATRIX,
NORMALIZED VARIABLES AND PHASE ADVANCES}

\vspace{-0.1cm}
Due to nonuniques of the diagonalizing matrix $T$ in the Williamson's theorem
the relation (\ref{matr_S2}) can be considered as a multi-valued function
which maps any particular matrix $\Sigma$ (and/or $W$) into some
subset of symplectic matrices. Let us select some single-valued branch
of this function which associates one, and only one, output to any
particular input. We will call any such branch as $T$-algorithm and the
examples of the $T$-algorithms will be given below.
So, let us fix some particular $T$-algorithm
and let us substitute the representation (\ref{matr_S2}) into the 
equation (\ref{intr_5}). Then, after some straightforward 
manipulations, we obtain

\vspace{-0.2cm}
\noindent
\begin{eqnarray}
\big(T(\tau) M(\tau) T^{-1}(0)\big)\cdot
\big(T(\tau) M(\tau) T^{-1}(0)\big)^{\top} = I,
\label{TP_2}
\end{eqnarray}

\vspace{-0.2cm}
\noindent
which means that the $2n \times 2n$ matrix

\vspace{-0.2cm}
\noindent
\begin{eqnarray}
R(\tau) \,=\, T(\tau)\, M(\tau)\, T^{-1}(0) 
\label{TP_3}
\end{eqnarray}

\vspace{-0.2cm}
\noindent
is orthosymplectic (i.e. orthogonal and symplectic simultaneously).
The equality (\ref{TP_3}), when written in the form

\vspace{-0.2cm}
\noindent
\begin{eqnarray}
M(\tau) \,=\, T^{-1}(\tau)\, R(\tau) \, T(0),
\label{TP_4}
\end{eqnarray}

\vspace{-0.2cm}
\noindent
gives us a (familiar in 1D) parametrization of the beam transfer matrix 
$M(\tau)$, and if we will introduce normalized variables $z_n$ by the equation

\vspace{-0.2cm}
\noindent
\begin{eqnarray}
z(s) \;=\; T^{-1}(s) \,z_n(s),
\label{TP_6}
\end{eqnarray}

\vspace{-0.2cm}
\noindent
then the dynamics in the normalized variables

\vspace{-0.2cm}
\noindent
\begin{eqnarray}
z_n(\tau) \,=\, R(\tau) \, z_n(0)
\label{TP_7}
\end{eqnarray}

\vspace{-0.2cm}
\noindent
is simply a rotation and the multidimensional Courant-Snyder invariant
(\ref{betas_77}) takes on the form $I_{cs}\,=\,z_n^{\top} \cdot z_n$.

Although the motion in the normalized variables (\ref{TP_7}) is not, in general, 
uncoupled motion, but from the point of view of the theory of beams with 
equal eigenemittances no additional simplifications are required,
because

\vspace{-0.2cm}
\noindent
\begin{eqnarray}
\Sigma_n
\,\stackrel{\mbox{\tiny def}}{=}\, 
\Big<z_n \cdot z_n^{\top}\Big>
\,=\,
\epsilon \,I
\label{TP_7_01}
\end{eqnarray}

\vspace{-0.2cm}
\noindent
is already a diagonal matrix.

Let us turn now our attention to the concept of phase advances, 
which does not have an unique choice even in the 1D case
(see, for example, discussion in ~\cite{Chao}). 
The phase advances are the quantities which should be associated with
the eigenvalues of the matrix $R$. This matrix
is orthosymplectic and can be partitioned into the form

\vspace{-0.2cm}
\noindent
\begin{eqnarray}
R \,=\,
\left(
\begin{array}{rr}
 C & S\\
-S & C
\end{array}
\right),
\label{QRTSYM_1}
\end{eqnarray}

\vspace{-0.2cm}
\noindent
where the $n \times n$ submatrices $C$ and $S$ satisfy

\vspace{-0.2cm}
\noindent
\begin{eqnarray}
C S^{\top} = S C^{\top},
\;\;\;\;\;
C C^{\top} + S S^{\top} = I.
\label{QRTSYM_2}
\end{eqnarray}

\vspace{-0.2cm}
\noindent
All eigenvalues of the matrix $R(\tau)$ lie on the unit circle in 
the complex plane, i.e. are of the form

\vspace{-0.2cm}
\noindent
\begin{eqnarray}
\exp(\pm i \mu_1(\tau)),\, \ldots,\, \exp(\pm i \mu_n(\tau)),
\label{QRTSYM_3}
\end{eqnarray}

\vspace{-0.2cm}
\noindent
and $\mu_m(\tau)$ are the quantities which we will
interpret as (fractional part of) phase advances.

If the beam transport in (\ref{intr_5}) is periodic
(i.e. if $\Sigma(\tau)=\Sigma(0)$ and, therefore, $T(\tau)=T(0)$), 
then the equality (\ref{TP_4}) tells us that the eigenvalues of the
matrix $R(\tau)$ are the same as the eigenvalues of the matrix $M(\tau)$.
It means that for the periodic beam transport the phase advances are uniquely
defined independently from any particular choice of the $T$-algorithm.
It is a very pleasant property, but it seems that it 
is the only property of the phase advances which does not depend 
from the choice of the $T$-algorithm. Let us, for illustration,
consider three 1D $T$-algorithms defined by the requirement for the
matrix $T$ in (\ref{matr_S2}) to be in one of the following special forms

\vspace{-0.1cm}
\noindent
\begin{eqnarray}
T_1 =
\left(
\begin{array}{cc}
 1 / \sqrt{\beta}     & 0\\
\alpha / \sqrt{\beta} & \sqrt{\beta}
\end{array}
\right),
\;
T_2 =
\left(
\begin{array}{cc}
\sqrt{\gamma} & \alpha / \sqrt{\gamma}\\
0             & 1 / \sqrt{\gamma}
\end{array}
\right),
\label{TP_5_1}
\end{eqnarray}

\vspace{-0.2cm}
\noindent
\begin{eqnarray}
T_3 \,=\,
\frac{1}{\sqrt{(\beta + 1) + (\gamma + 1)}}
\left(
\begin{array}{cc}
\gamma + 1 & \alpha\\
\alpha     & \beta + 1
\end{array}
\right).
\label{TP_5_3}
\end{eqnarray}

\vspace{-0.1cm}
\noindent
Note that here $T_1$ is the original Courant-Snyder choice and
$T_3$ is symmetric (and symplectic) positive definite 
square root of the matrix $W^{-1}$.
In all these cases the
matrix $R(\tau)=R(\mu_m(\tau))\,$ in (\ref{TP_4}) is given by

\vspace{-0.15cm}
\noindent
\begin{eqnarray}
R(\mu_m(\tau)) =
\left(
\begin{array}{rr}
 \cos(\mu_m(\tau)) & \sin(\mu_m(\tau))\\
-\sin(\mu_m(\tau)) & \cos(\mu_m(\tau))
\end{array}
\right),
\label{TP_5_2}
\end{eqnarray}

\vspace{-0.15cm}
\noindent
and, in order to see more clearly the difference between the behavior 
of the phase advances $\mu_m$, let us assume that the dynamics is derived from
the Hamiltonian

\vspace{-0.15cm}
\noindent
\begin{eqnarray}
H(\tau) \,=\,(1 \,/\, 2) \cdot
\big(\, p_1^2 \,+\, k(\tau) \,q_1^2\,\big).
\label{TP_5_4}
\end{eqnarray}

\vspace{-0.15cm}
\noindent
Then the phase advances $\mu_m$ obey the equations

\vspace{-0.15cm}
\noindent
\begin{eqnarray}
\frac{d \mu_1}{d \tau} \,=\,\frac{1}{\beta},
\;\;\;\;\;
\frac{d \mu_2}{d \tau} \,=\,\frac{k}{\gamma},
\label{TP_5_5_1}
\end{eqnarray}

\vspace{-0.4cm}
\noindent
\begin{eqnarray}
\frac{d \mu_3}{d \tau} \,=\,\frac{k\, (\beta \,+\, 1) \,+\, (\gamma \,+\, 1)}
{(\beta \,+\, 1) \,+\, (\gamma \,+\, 1)}.
\label{TP_5_6}
\end{eqnarray}

\vspace{-0.15cm}
\noindent
One sees that while $\mu_2$ stays constant in the
drift spaces, $\mu_1$ and $\mu_3$ change; 
while $\mu_1$ changes monotonously, $\mu_2$ and $\mu_3$
can be locally increasing and decreasing; and etc. 
Besides that, let us note that the multiplication of any matrix
$T_m$ from the left by a constant 
rotation matrix does not change the corresponding phase advance.
For example, for an arbitrary constant angle $\psi$ 
the $T$-algorithm associated with the matrix 

\vspace{-0.2cm}
\noindent
\begin{eqnarray}
\left(
\begin{array}{rr}
 \cos(\psi) & \sin(\psi)\\
-\sin(\psi) & \cos(\psi)
\end{array}
\right) \cdot T_1
\label{TP_5_1_01}
\end{eqnarray}

\vspace{-0.15cm}
\noindent
will produce the same phase advance as in the original Courant-Snyder case,
but, in general, the triangular form of the transition to the normalized variables
(\ref{TP_6}), which is also important for 1D theory,
will be lost.
So, it seems that the correct way to extend the Courant-Snyder choice
of the 1D phase advance to the multidimensional case without losing
what else important comes through the direct generalization of the 1D 
$T$-algorithm defined by the matrix $T_1$ to many dimensions.

The matrix $T_1$ is a lower triangular matrix with positive diagonal elements
and as its multidimensional analog we will take the 
lower block triangular symplectic matrix

\noindent
\begin{eqnarray}
T \,=\,
\left(
\begin{array}{cl}
w^{-1}   & 0\\
u w^{-1} & w^{\top}
\end{array}
\right),
\label{A2_3}
\end{eqnarray}

\vspace{-0.2cm}
\noindent
where $u$ and $w$ are, respectively, a symmetric and 
an invertible $n \times n$ matrix. 
Note that any lower block triangular symplectic matrix
can be represented in this form with the proper choice
of the matrices $u$ and $w$.

\vspace{-0.15cm}
\begin{proposition}

The beam matrix $\Sigma$ can be diagonalized by congruence using 
symplectic block triangular transformation of the form (\ref{A2_3}) 
if and only if

\vspace{-0.2cm}
\noindent
\begin{eqnarray}
(\Sigma J)_{qp}^2 \;=\;
\Sigma_{qq}\,\Sigma_{qp}^{\top}
\;-\;
\Sigma_{qp}\,\Sigma_{qq} \;=\;0,
\label{A2_7}
\end{eqnarray}

\vspace{-0.2cm}
\noindent
and the condition (\ref{A2_7}) is invariant under linear symplectic 
transport (\ref{intr_5}) of the beam matrix $\Sigma$ if and only 
if the matrix $\Sigma$ is the matrix with equal eigenemittances.

\end{proposition}

\vspace{-0.15cm}
\noindent
This proposition tells us that though not only
$\Sigma$ matrices with equal eigenemittances can be diagonalized
by the matrix (\ref{A2_3}), the consistent theory of diagonalization 
by the lower block triangular symplectic matrices can be created only for
the beams with equal eigenemittances.

Substituting representation (\ref{A2_3}) into relation (\ref{matr_S2})
we obtain the following equations for the determination of
the matrices $w$ and $u$ as functions of the given matrix Twiss 
parameters $\beta$, $\alpha$ and $\gamma$

\vspace{-0.15cm}
\noindent
\begin{eqnarray}
\beta = w \,w^{\top},
\label{TP_1_1}
\end{eqnarray}

\vspace{-0.4cm}
\noindent
\begin{eqnarray}
\alpha = w \,u\, w^{-1},
\label{TP_1_2}
\end{eqnarray}

\vspace{-0.4cm}
\noindent
\begin{eqnarray}
\gamma = w^{-\top}(I + u^2) \,w^{-1}.  
\label{TP_1_3}
\end{eqnarray}

\vspace{-0.15cm}
\noindent
The general solution of the equation (\ref{TP_1_1}) can be written
as $w =\hat{w}\,r$, where $\hat{w}$ is any particular solution of
this equation and $r$ is an arbitrary $n \times n$ orthogonal matrix.
When some solution of the equation (\ref{TP_1_1}) is chosen, then
the matrix $u$ is uniquely defined from the equation (\ref{TP_1_2})
and is symmetric,
and the equation (\ref{TP_1_3}) is satisfied automatically due to
relation (\ref{betas_4}).
Thus the only remaining uncertainty in the choice of the $T$-algorithm
lies in the nonuniques of the solution of equation (\ref{TP_1_1}).
Unfortunately, this  uncertainty cannot be resolved simply by the requirement
to recover the Courant-Snyder choice in the limit of multidimensional
uncoupled beams. 
For example, both, $w$ taken as unique Cholesky factor of the matrix $\beta$ and
$w= \beta^{1/2}$, where  $\beta^{1/2}$ denotes an unique positive definite 
symmetric square root of the matrix $\beta$, will satisfy this requirement.
But if we will take into account that the Courant-Snyder matrix $T_1$ has
positive diagonal elements and will reformulate this property in the form
that $1 \times 1$ diagonal submatrices of the matrix $T_1$ are symmetric
positive definite matrices, then the choice for $w$ becomes unique.
One takes $w = \beta^{1/2}$ and obtains for the matrix $T$ 
the final (and familiar from 1D theory) form

\vspace{-0.2cm}
\noindent
\begin{eqnarray}
T \,=\,
\left(
\begin{array}{cc}
\beta^{-1/2}           & 0\\
\beta^{-1/2} \, \alpha & \beta^{1/2}
\end{array}
\right).
\label{FC_1}
\end{eqnarray}

\end{document}